# Magnetic Particles for Multidimensional In-vitro Bioanalysis

Gungun Lin[1,2*]

Email: gungun.lin@uts.edu.au

1. Institute for Biomedical Materials and Devices, Faculty of Science, The University of Technology Sydney, Ultimo, New South Wales 2007, Australia.
2. ARC Research Hub for Integrated Device for End-User Analysis at Low Levels, Faculty of Science, University of Technology Sydney, Sydney, NSW 2007, Australia

**Abstract:**

Multidimensional or multiplex bioanalysis represents a crucial approach to improve diagnostic precision, increase assay throughput and advance fundamental discoveries in analytical industry, life science and nanomedicine. Along this line, bio-interfacing magnetic particles have been playing an important role. Fully exploiting the properties of magnetic particles is the key to tailoring recent technology development for better translational outcomes. In this mini-review, typical magneto-physical dimensions of magnetic particles are introduced. Recent progress of implementing these dimensions with advanced sensor and actuator technologies in multiplex bioanalysis is discussed. Outlooks on potential biomedical applications and challenges are provided.

## 1. Introduction

A substantial part of research in analytical chemistry, biosensing and nanomedicine requires the identification of intercorrelated molecular or cellular parameters to shape translational outcomes. For instance, biomarkers discovered through such a process could be promising as surrogate clinical endpoints. Following this trend, there has been a surge in the pursuit of multiplex bioanalytical capability, i.e. the ability to profile multiple variants at the same time or in a common biological specimen. However, challenges are as significant as regard to the lack of efficient tools for high-dimensional analysis. While innovations in functional materials can impact many advanced sensors and actuators technologies, progress made should ultimately expand the capacity in both fundamental research and applications.

Owing to the advances in synthesis chemistry [1,2], magnetic particles (here refer to a class of functional micro- and nanostructures with sizes ranging from 10 nm to 100 um capable of reacting to a magnetic field) are playing an increasing role in the analytical and biomedical fields for their unique magneto-physical properties. For example, the magneto-mechanical property of magnetic particles has enabled biotechnology sectors to isolate, enrich and sort biological analytes. There has been significant potential to explore other intrinsic magneto-physical properties, such as non-linear magnetisation, magnetic flux, relaxation and thermal effects of magnetic particles. The fascinating properties of magnetic particles and their compatibility to interface with various biological entities have fuelled opportunities in the past decades. The trend is boosted by the achievements of advanced magnetic field sensor technologies capable of detecting magnetic field down to femto-tesla.

As of to date, numerous review articles have been devoted to the physics of magnetic field sensor [3–5], specific aspects of magnetic nanoparticles, including their dynamic resonance for biosensing [6], representative medical applications [7], or the cooperation with magnetic field sensors for biomolecular analysis [8–10]. The scope of this mini-review article on the exploration of the intrinsic properties of magnetic particles for in-vitro multidimensional analysis of different scales of biological analytes (Figure 1). In the first section, representative magneto-physical properties of magnetic particles are discussed. For the second section, current implementation in sensing and actuating techniques and promising applications in multidimensional bioanalysis, including planar and suspension assays are provided. Finally, outlooks on future directions and opportunities are summarized.

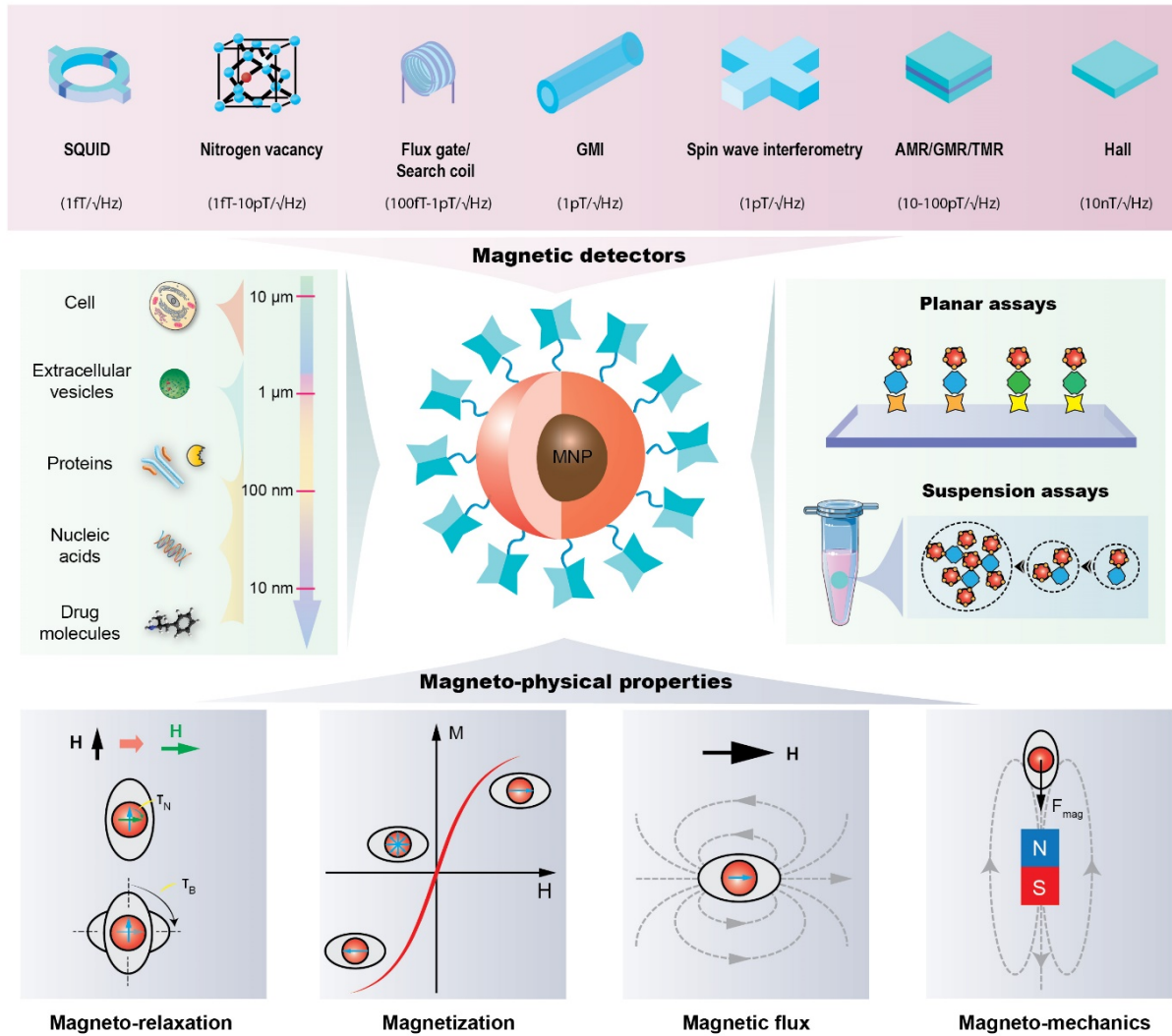

Figure 1. Overview of exploiting magnetic particles for in-vitro multidimensional bioanalysis. (A) Magnetic detectors with different sensitivities. (B) Bio-functionalized magnetic particles can interface with multiscale biological entities, such as drug molecules, proteins, nucleic acid, extracellular vesicles and cells. (C) Two representative magnetic assay formats: planar assay and suspension assay. (D) Representative magneto-physical properties of magnetic particles explored for biosensing and analysis.

## 2. Fundamentals of magneto-physical properties and technological implementations

### 2.1. Non-linear magnetization

B-H curves are generally used to describe the behavior of a magnetic material, which provide B, magnetic flux, as a function of H, namely, external magnetic field, in different directions. The susceptibility (denoted by χ) of a material can be defined by the tendency of increase or decrease of the resultant magnetic field inside the material comparing with the applied magnetic field. Many materials in the bioassay volume, such as glass slides, plastic tubes, water, cells, molecules or polymer matrices are paramagnetic (χ >0) or diamagnetic (χ<0). Their B-H curves are linear with the change of magnetic field. In contrast, ferromagnetic materials such as cobalt, iron and rare earth alloys exhibit large magnetic susceptibilities (χ >>0). Their magnetization remains after the removal of magnetic field. Superparamagnetic particles with size about 10 nm do not carry remanence upon field removal and possess a kind of strong and non-linear magnetization behavior. This property distinguishes them from other materials, in particular, those biological species, endowing the particles one of the most promising labels of biological cells or molecules.

The nonlinear magnetization property of magnetic particles has been successfully implemented in a "Frequency Mixing (FM)" sensing technique [11] and magnetic particle imaging (MPI)[12] based on superparamagnetic nanoparticle labels. The 'FM' technique can be used to extract weak signals from interfaces against large background signals. It functions typically by exciting magnetic particles of nonlinear magnetic characteristics at two different frequencies, i.e., $f_1$ and $f_2$. The resultant response signal, i.e., frequency-modulated magnetic flux, from the particles can manifest itself as a linear combination of $f_1$ and $f_2$, $f_{new} = mf_1 \pm nf_2$, where $m$ and $n$ are natural numbers, making it unique for particles with nonlinear magnetization characteristics. The "magnetic frequency mixing" is superior to susceptometry for identifying the superparamagnets, as paramagnetic water can carry more susceptometric signals than the superparamagnets. The amplitude of the frequency response of magnetic beads can be related to the amounts of magnetic materials, suggesting its use for generic magnetic quantification [13]. In contrast, the phase response is invariant with respect to the change of the concentration of magnetic beads. These parameters offer the opportunity to differentiate a mixture of two types of magnetic particles using the signal amplitude and phase of frequency response. In MPI, a "Field Free Point (FFP)" is scanned swiftly over a sample to produce a tomographic image. The superparamagnets are saturated at every point except for this FFP. Particles at this point reverses their magnetization to produce an MPI signal. Since the tissue samples are transparent to the magnetic field, only superparamagnetic particles injected in the tissues are detected.

## 2.2. Magnetic relaxation

Magnetic relaxation describes the process during which a magnetic system establishes an equilibrium or steady state condition under a pulsed magnetic field. Three types of magnetic relaxation have been majorly explored for in-vitro bioanalysis: Brownian relaxation (BR) [14–16], Néel relaxation (NR) [17,18] and spin-spin relaxation (T2) [19–23].

BR refers to the physical rotation of particles with a hydrodynamic volume of $V_H$ that are swimming in a solution. The relaxation time is expressed as:

$$\tau_B = \frac{3\eta V_H}{k_B T}$$

where $\eta$ is the viscosity of the solution, $V_H$ is hydrodynamic volume of the particle, $k_B$ is Boltzmann constant, and T is absolute temperature in Kelvin.

NR corresponds to the reorientation of the magnetic moment of particles. Its relaxation time can be

given by:

$$\tau_N = \tau_0 \exp\left(\frac{KV_m}{k_BT}\right)$$

where the time constant $\tau_0 = 10^{-9}$, K is the anisotropy constant of magnetic particles, $V_m$ is the volume of the magnetic particle. For magnetic nanoparticles, the time for Neel relaxation is in the order of milliseconds to seconds, while for Brownian relaxation, it can take place in microseconds.

Both BR and NR have been implemented in relaxometry with highly sensitive magnetometers such as superconducting quantum interference devices (SQUID), low-noise fluxgates, Hall sensors [24,25] and inductive coils [26,27] and giant magnetoresistance (GMR) sensors [18] as the detectors. The pulsed magnetic field for magnetic excitation can be provided by an ultrafast electromagnet. The time-course magnetization of magnetic particles can be strongly dependent on the duration and strength of the applied pulsed magnetic field, requesting the magnetic field to be uniform for saturating the particles. The probing of the magnetic relaxation can take place after drying the particles. Hence, detection with magnetic relaxation can bypass the process of monitoring the real-time binding curves in wet samples, or the tricky step to balance the background signal level of reference sensors and probe sensors, which is required for static-field magnetometry. In this respect, the technique is arguably simpler than static-field magnetometry.

Spin–spin relaxation refers to the process by which the transverse component of the magnetization experiences an exponential decay or dephasing towards its equilibrium state. It can be measured by the spin–spin relaxation time, known as T2, a value characterizing the signal decay.

$$M_{xy} = M_{xy}(0)e^{-t/T_2}$$

Spin-lattice relaxation and its time, T1, characterizes the process of the longitudinal magnetization recovering to its initial state. Both T1 and T2 can be used in nuclear magnetic resonance (NMR) and magnetic resonance imaging (MRI), while in biomolecular assays, T2 is more broadly used. T1 can be altered by the dipolar coupling between the proton moments with surroundings. T2 is dependent on the molecular structures and the amounts of hydrogens in the tissues (e.g. $6.6 \times 10^{19}$ protons per $mm^3$ of water, water-based tissues: 40-200ms, fat-based tissues: 10-100 ms;)[28], and can be influenced by the local inhomogeneity of applied longitudinal field. In this respect, paramagnetic complexes (e.g. transition and lanthanide metal ions with unpaired electrons) and superparamagnetic nanoparticles have been used to enhance the T1 and T2 image contrast, respectively.

## 2.3. Magnetic flux

As the most intrinsic properties of magnetic particles, magnetic flux is another signature to distinguish magnetic labels from other non-magnetic biological cells or molecules that do not emit magnetic flux. However, certain antiferromagnetically-coupled structures or those carrying vortex domains are designed not to emit magnetic flux. Magnetic field sensors can be implemented to detect the magnetic flux of particles in the time domain that gives rise to relaxometry, or in the spatial domain resulting in static field magnetometry (most often based on magnetoresistance sensors and Hall sensors).

The magnetic flux of a magnetic particle is frequently estimated by assuming the particle as a magnetic dipole. The magnetic dipole field is expressed as[29]:

$$B(r) = \frac{\mu_0}{4\pi}\left[\frac{3r(m \cdot r)}{r^5} - \frac{m}{r^3}\right],$$

where r is the vector connecting the dipole center and the position of the magnetic field, $\mu_0$ is the vacuum susceptibility, and m is the magnetic moment. This equation suggests the inherent 3D nature of magnetic flux that can decay rapidly with the increase of the separation distance. Sectioning the magnetic flux in various time and space can also reveal distinctive patterns. All the features have posed a significant detection challenge.

Magnetic particles with differentiable magnetic moments can bring in a so-called concept of magnetically-defined 'barcodes' (with total coding capacity of 6-7) [30]. The spatial distribution of magnetic flux can be engineered to expand the coding capacity, by artfully fabricating barcode-like ferromagnetic particles comprising multiple magnetic segments/compartments with each emitting magnetic flux along a particular direction to represent digital "1" or "0" [31]. The magnetization direction of the ferromagnetic elements can be individually set by an external magnetic field because of their different coercivities. However, the information could be erased by magnetic field that prohibits a key advantage of magnetic particles for, e.g. isolating biological samples. The residual magnetic flux could further lead to interparticle dipole-dipole interactions, causing agglutinations. An alternative approach can be the coding of information into the architecture of microbeads consisting of superparamagnetic nanoparticles [29]. Using colloidal assembly, magnetic coupling between the magnetic colloidal compartments with tailorable number and size ratio can give rise to tailorable magnetic flux patterns. The tailorable coupling can result in tunable critical rotation frequency of magnetic particles [32], which is defined as the frequency at which the particles transition from synchronous rotation to asynchronous rotation". By using a global control signal, namely, a rotating magnetic field, this method shows potential to merge the magnetic manipulation with the coding and decoding process.

## 2.4. Magneto-mechanics

Depending on the size of magnetic particles, magneto-mechanics can involve ferrohydrodynamics (FH), magnetorheology (MR) and magnetophoresis (MP) [33]. FH and MR apply to a class of high-concentration magnetic nanoparticles with sizes ranging from 10 nm to 1 micrometer which can form a bulk continuum of magneto-fluids. The viscosity of the fluids can be altered by applying a magnetic field. In contrast, magnetophoresis deals with the motion of discrete magnetic entities towards high magnetic field gradient. The magnetic entities can be single magnetic particles or minute amounts of magnetic particles conjugated to biological analytes.

The force, $F_m$, acting on a magnetically conjugated entity can be expressed as [34]:

$$\boldsymbol{F}_m = \frac{V(\chi_p - \chi_f)}{\mu_0}(\mathbf{B} \cdot \nabla)\boldsymbol{B}$$

Where V is the volume of the magnetic entity, B is the magnetic field, $\chi_p$ and $\chi_f$ is the susceptibility of the magnetic entity and surrounding fluid, respectively, $\mu_0$ is the permeability. Permanent magnets, electrical coils or patterned ferromagnetic microstructures can be designed to locally alter the magnetic field gradient experienced by the magnetic entities to realize magnetophoretic focusing, levitation and sorting. As the above equation implies, the magnetic force is reliant on the susceptibility contrast between the particles and surrounding environments. Diamagnetic bioparticles, such as cells [35,36] and extracellular vesicles [37] can be suspended in ferrofluid/paramagnetic medium and negative magnetophoresis can occur to isolate these non-magnetic particles. An emerging trend along this direction has witnessed the combination of various forces such as viscoelastic forces with magnetophoresis to improve the sorting efficiency of nonmagnetic particles [38].

## 2.5. Magneto-thermal property

Magnetic particles can release heat when exposed to a magnetic field alternating at high frequencies in the order of kilohertz and megahertz. The heat is generated because of the hysteretic and relaxation losses. In this respect, ferromagnetic particles with multidomain structures may have different magnetothermal properties from superparamagnetic particles possessing single-domain structures. For biomedical applications, the latter has broader interest due to its small size allowing them to be internalised into cells and tissues or conjugated to biomolecules with minimum steric hinderance. Single-domain superparamagnetic nanoparticles do not have hysteretic magnetization properties, thus at low field frequencies, they don't exhibit magnetic heating effects. However, at higher frequencies, the Neel relaxation of the particles can become hysteretic. This corresponds to frequencies closed to the relaxation frequency. Alternatively, Brownian relaxation of the particles can also generate heat due to the shear stress between the particles and surrounding environments. As noted previously, both the Neel and Brownian relaxation can be controlled by the intensity of the applied magnetic field. Neel relaxation is related to the magnetic anisotropy of the whole system. By adding a static magnetic field to the existing alternating field [39] or tailoring the magnetic anisotropy constants through dipolar coupling, i.e. ordering of nanoparticles [40] or spin-orbit coupling, e.g. by doping transitional metal ions (e.g. Co, Mn and Ni) into iron oxides [41] has proven to be effective to maximise heat dissipation. Magnetothermal multiplexing can enable independent control over multiple processes for local stimulation of tissues and triggering of drug release using combinations of frequency and amplitude controls of magnetic fields.

## 3. Current Implementation of Magnetic Particles in Bioanalysis

### 3.1. In-vitro biomolecular analysis

Quantification of multiple biomolecular analytes, such as DNA and proteins using the multidimensional properties of magnetic particles has been applied for disease diagnosis, pathogen testing, environmental monitoring and drug screening. For instance, femtomolar ($10^{-15}$) proteins labelled by magnetic nanoparticles have been detectable with magnetoresistance-based nanosensors [42]. In this scenario, 64 magnetoresistance nanosensors in an 8x8 array were fabricated on a 1.2 cm x 1 cm silicon chip. A sandwich assay format was adopted, in which the target antigen was sandwiched between two antibodies, one attached to the sensor surface, the other tagged with superparamagnetic nanoparticles. The sensing scheme was proven insensitive to PH, optical activity and turbidity, ionic strength and temperature that could be varied in real-world biological samples. This was different from other sensing schemes, such as nanowires, in which significant signal fluctuation can be caused by only 0.5 PH change, or microcantilevers, for which a 0.5-degree change can result in substantial beam deflection. Multianalyte and multiprobe assay can be achieved with this format by functionalizing different regions of the sensor arrays with capture antibodies [43].

Real-time binding of the antigen to the surface antibodies brings the magnetic nanoparticle tags to the proximity of sensors, causing graduate increase of the magnetic flux detected by the sensors. The real-time biding signal curve can be used to quantify the interaction between labelled macromolecules and targets immobilized on a sensor surface [44]. This can complement the gap in understanding the binding kinetics between labels and targets in heterogeneous assays. Compared with surface plasma resonance (SPR), the standard for monitoring protein binding interactions, the sensing scheme with magnetic nanoparticle tags can enable parallel analysis of hundreds of thousands of reactions at a dynamic range of > 6 log on a single chip with an area of 1 cm$^2$. In stationary solution, a two-compartment model can be established for estimating the kinetic parameters, applicable to DNA hybridization assays as well [45]. More recently, Wang et al. extended the scheme for probing low affinity immune check points receptors and their ligands targeted in immunotherapies, such as PD-L1/PD-L2 interactions [46]. The elevated detection sensitivity over SPR eases the intensive use of proteins of low affinity interactions. Implementing the sensors in a microfluidic chip to flow magnetic particle-conjugated proteins over the sensor surface can also reduce he two-compartment model into a simpler Langmuir isothermal.

## 3.2. Phenotyping of cells

Profiling cellular targets allow for identifying tumor types, monitoring tumor burdens and therapies. For instance, the availability of serial tumor tissue has been limited because traditional biopsies are procedural time-consuming, costly and invasive. The small amounts of cells or tissues obtained by fine-needle aspirates are not sufficient for conventional molecular profiling approaches, such as immunohistochemistry, proteomics analyses or flow cytometry, which often require considerable quantities of cell samples. More sensitive technologies to detect limited quantities of lesion cells are of intense interest. A micro-NMR technology by measuring the transverse relaxation rate ($T_2$) of magnetic nanoparticles attached to the cells can be used for multiplex molecular biomarkers profiling of cells obtained by fine-needle aspirates from suspected lesions from patients [21]. Considerable and unexpected heterogeneity on biomarker expression levels emphasizes the necessity of multiplex analysis and has crucial implication for molecular diagnostics and therapeutic drug targeting.

Blood is inherently a mixture of many abundant cell types, such as erythrocytes and monocytes and rare cells, such as circulating tumour cells. Adapting magnetic assays for flow cytometry can enable rare cell analysis (<100 cells/mL whole blood), including enumeration and profiling of specific biomarkers at the single cell level. Using magnetic nanoparticles to selectively target the biomarkers of rare cells, such as EpCAM, HER2/neu and EGFR, allows differentiating different cell lines. Issadore et al. presented a micro-Hall sensor chip to detect the magnetic flux of rare cells labelled by functionalised magnetic nanoparticles [30]. The abundance of biomarkers on the cell membrane is proportional to the loading of magnetic nanoparticles and thus the total magnetic flux intensity. Multiplex detection of individual cells can be achieved by using manganese-doped ferrite ($MnFe_2O_4$) superparamagnetic nanoparticles of different diameters (i.e. 10, 12, and 16 nm), each with different magnetization curves. Each type of magnetic nanoparticles is functionalised with receptors to simultaneously target the EpCAM, HER2/neu and EGFR biomarkers on the same cells. By magnetizing the cells using different external field strength, the relative abundance of biomarkers on the cells can be calculated based on the known magnetization curves of different magnetic nanoparticles.

The differentiable amounts of magnetic particles on the cells also enables magnetically activated sorting of the cells based on the abundance of expressed biomarkers. The subsequent molecular profiling can be virtually done with dye staining for deep characterization. Magnetic ranking cytometers based on multiplex magnetic sorting can be used for profiling rare cells (10 to thousand cells) and for large number of cells (up to tens of millions), respectively [47]. The systems can be generalised to be relying on rationally-designed microfluidic channel layout and local tailored magnetic field gradient produced by some micro-structured ferromagnetic elements purposely coated at the channel bottom. The magnetophoretic sorting and magnetic trapping allow for capturing rare cells or isolating a large population of cells conjugated with different amounts of magnetic nanoparticles into different outlets. Apart from the use for determination of cell surface proteins, the incorporation of magnetophoretic cell sorting in microfluidics can be applied for studying tumour cell intravasation and extravasation [48], transcriptomics analysis of single cells [49], identifying rare human pluripotent stem cells [50] and antibiotic resistant bacteria [51], as well as phenotypical CRISPR screens [52].

## 4. Outlook: Future Directions and Applications

**At the molecular level**, in-vitro analysis could go beyond the single physical dimension of magnetic particles. Suspension assays utilizing beads as molecular carriers are generally advantageous for its fast reaction kinetics compared with planar arrays [53]. Some pioneering works show the promise of using magnetic flux for coding magnetic beads [54]. Such scheme is compatible with flow cytometry, making this method a potential high-throughput solution. Nonetheless, a holistic streamlined magnetic assay solution based on magnetic coding and reporting is not yet reported, as both coding and reporting using the same magnetic dimension could interfere with each other. This is in stark contrast with optical multiplexing, where each color can be separated by optical filters [55]. Hence, effectively combining mutually non-interference magneto-physical dimensions for coding and reporting are desired. Moreover, single molecule detection could be feasible with the advance of ultrasensitive nanoscale sensors such as diamond nitrogen-vacancy center magnetometers [56] and the synthesis chemistry of magnetic nanoparticles with high saturation magnetization [57]. Most previous studies have focused on the 'passive' sensing abilities of magnetic particles for in-vitro biomolecular assays. The 'active' actuating or heating aspects of magnetic particles could open up new opportunities. For instance, the combination of time-domain measurement, such as magnetic relaxation and magneto-mechanics should enable to probe the force-response molecular relaxation dynamics, such as DNAs folding-unfolding by integrating magnetic tweezers with magnetic sensors.

**At the living cell level**, growing efforts are devoted towards single cell analysis in order to decode the inherent complexities of cell heterogeneity, where magnetic particles could play an important role. For instance, DNA barcoded hydrogel particles are co-loaded with single cells in droplets to profile single cell omics [58]. It could be natural to utilize magnetically-coded particles to streamline the process of analyte harvesting and single cell barcoding. There are other possibilities beside the pure quantification of biomolecular analytes released by living cells. The magneto-mechanical and -thermal effects of particles can allow phenotyping cells based on their response to external stimuli, such as heat or mechanical force induced movement of cells, secretion, enzyme production and gene expression of single cells [59]. Targeted delivery of magnetic nanoparticles into specific locations in a cell, such as organelles, could enable in-situ intercellular assays. Micro-viscometers based on asynchronous rotation of magnetic microbeads have been reported [60]. Using magnetic nanoparticles can allow intracellular viscosity sensing. Magnetic nanoparticles could be conjugated with fluorescent dyes or other luminescent nanoparticles for motion tracking [61]. All these new dimensions could be added to the existing panel of parametric analysis for disease stratification and could innovate therapeutic solutions.

**At the tissue level**, sensing or actuation of magnetic particles may create a myriad of opportunities although challenging. Optical evaluation of multicellular samples or tissue-like structures is challenging because of light penetrating and scattering issues. Magnetic particles modified with hydrogel surfaces can be used as carriers of 3D cellular constructs, the actuation of which could be navigated by external magnetic field for high-quality optical evaluation of tissue-like samples [32]. Assembly of magnetic particle building blocks allows constructing smart micro-robots for tissue assembly [62]. It is expected that, in the longer term, magnetically-controlled micro-robots formed by a swarm of magnetic particles with programmable motility and shape-morphing ability could be used for tissue repair and surgery [63]. Magnetic beads decorated with soft and biocompatible polymers may interface with tissue or tissue-like multicellular structures. Magnetically-responsive beads can be utilized to apply forces to cells to study mechano-transduction, as in magnetic twisting cytometry [64]. When engineered with biocompatible interface, they can be placed inside tissues to apply forces and measure tissue associated forces at the supracellular length scale [65]. Furthermore, emerging techniques are seeking to perform cellular analysis while keeping the spatial information of cells relative to other cells within a tissue, which is vital for revealing the correction between cell microenvironments and their phenotypes [66]. This requires high-capacity cell barcodes to register the spatial location of cells that could be subsequently profiled by downstream high-throughput single cell analysis techniques. The use of magnetically-barcoded particles may facilitate cell barcoding, dissociation, collection and identification.

In this mini-review, I have been able to highlight a limited number of key bioanalysis applications of magnetic particles that have leveraged the essential multidimensional magneto-physical properties. They have enabled 'passive' sensing in cooperation with magnetic field sensors, or the 'active' actuating of biological cellular or molecular species using magnetic field. It is envisioned that the coalescence of multiple magnetic dimensions of magnetic particles may go well beyond the simple addition effect and can significantly increase the depth of basic bio-discoveries and the breadth of clinical applications in the long term.

**Acknowledgements**

The author acknowledges the financial support from National Health and Medical Research Council
Fellowship (GNT: 1160635), the ARC Industry Transformational Research Hub Scheme (grant
IH150100028).